\documentclass[12pt]{article}
\usepackage{amsmath}
\usepackage{amssymb}
\usepackage{graphicx}
\usepackage{cite}
\usepackage{color}
\usepackage{pstricks,calc,pst-node,graphics}
\usepackage{amsgen,amsmath,amssymb,color}

\usepackage[labelfont=bf,labelsep=period,justification=raggedright]{caption}

\makeatletter
\renewcommand{\@biblabel}[1]{\quad#1.}
\makeatother

\date{}

\newcommand{\tocompletesameline}[1]{
{\textcolor[rgb]{0.50,0.00,0.00}{\textbf{(to complete) }}}}

\newcommand{\E}{\mathrm{I\!E}}

   \newcommand{\newpar}{{\vspace{0.3cm} \noindent}}

\newcommand{\indep}{\mbox{$\perp\!\!\!\perp$}}

\psset{arrows=->,fillcolor=white,fillstyle=solid}
\newcommand{\Show}[1]{\psovalbox{#1}}
\newcommand{\Square}[1]{\psframebox{#1}}

 \definecolor{pink}{rgb}{1,
 .75, .8} \definecolor{emcolor}{rgb}{1,0,0}

\begin{document}

{\huge
{\textcolor[rgb]{0.00,0.00,0.50}{\textbf{Direct
genetic effects and their estimation from matched
case-control data }
}}}

\vspace{0.7cm}

\large

\begin{flushleft}

Carlo Berzuini$^{1}$, Stijn Vansteelandt$^{2}$, Luisa Foco$^{3}$,
Roberta Pastorino$^{3}$, Luisa Bernardinelli$^{3,1}$

\vspace{0.7cm}

\begin{description}
\item[1] Statistical Laboratory, Centre for Mathematical Sciences, University of Cambridge, United Kingdom
\item[2] Department of Applied Mathematics and Computer Science,\\
Ghent University, Belgium
\item[3] Department of Applied Health Sciences, University
of Pavia, Italy
\end{description}

\vspace{0.5cm}

\end{flushleft}

\normalsize

\noindent {\small In genetic association studies, a single
marker is often associated with multiple, correlated
phenotypes (e.g., obesity and cardiovascular disease,
or nicotine dependence and lung cancer). A pervasive
question is then whether that marker has independent
effects on all phenotypes. In this article, we
address this question by assessing whether there
is a \textit{direct} genetic effect on one
phenotype that is \textit{not} mediated through
the other phenotypes. In particular, we
investigate how to identify and estimate
such direct genetic effects on the basis of
(matched) case-control data. We discuss
conditions under which such effects are
identifiable from the available (matched)
case-control data. We find that direct
genetic effects are sometimes estimable via
standard regression methods, and sometimes
via a more general G-estimation method,
which has previously been proposed for
random samples and unmatched case-control studies~\cite{Vansteelandt2009,
Vansteelandt2009-genepi}
and is here extended to matched case-control studies.
The results are used to assess whether the FTO gene
is associated with myocardial infarction other
than via an effect on obesity.}

\section{\textcolor[rgb]{0.00,0.00,0.50}{Introduction}}

Associations of a genetic variant with a primary
phenotype can be difficult to interpret when one
considers the likely presence of correlated
phenotypes. The genetic association may then
be the indirect result of genetic effects
on a correlated phenotype, which subsequently
affect the primary phenotype. For instance,
Chanock and Hunter \cite{Chanock} discuss
three genetic association studies which
identified an association between a genetic
variation on chromosome 15 and the risk of
lung cancer, but the studies disagree on
whether the link is direct or mediated through
nicotine dependence. Addressing this question
may be important
to a better understanding of the underlying
causal mechanism.
This article addresses the general problem of
inferring the direct effect of a marker $X$ on
a trait $Y$ (e.g., lung cancer), controlling
for a correlated trait $M$ (e.g., nicotine
dependence), which we will refer to as a
{\em mediating} variable or mediator.

\newpar Vansteelandt {\em et al.}~\cite{Vansteelandt2009-genepi}
consider this problem in the context of prospective studies
of genetic association. Motivated by
the frequent use of ascertained samples
in those studies, in this paper we extend
the method to {\em matched} case-control
studies. We show that case-control sampling
seriously complicates the identification of
direct genetic effects. Progress
can be made within certain classes of statistical
models and under specific no unmeasured confounding
assumptions. In particular, we find that,
under very restrictive
conditions, direct
effects are estimable from
case-control data by using standard
regression methods, and that they are
estimable under more lenient conditions
by using special $G$-estimation methods~\cite{Vansteelandt2009},
which we here extend to matched case-control
data. In this paper, the required conditions
for estimability are unambiguously expressed
as conditional independence relationships
between problem variables, which we can check
on a
causal diagram~\cite{pearlbook,Dawid2002,
Greenland2000}.
We illustrate the method with the aid of a motivating
study, in which we use matched case-control data
to assess
whether variation in the chromosomal region
of the FTO gene causally affects
susceptibility to myocardial infarction other
than via an increase in body mass.

\section{\textcolor[rgb]{0.00,0.00,0.50}{Motivating study}}
\label{Motivating study}

FTO is a large gene on chromosome 16, that is highly expressed in the
hypothalamic nuclei that control eating behaviour in
mice~\cite{Dina2007}. The first intron of FTO harbours
the single nucleotide  polymorphism (SNP) rs9939609,
associated with body mass~\cite{Nature2007} and
myocardial infarction~\cite{Scott2007,Zeggini2007,Frayling2007,
Scuteri2007,Dina2007,Andreasen2008,Peeters2008}.
A simple, tentative, interpretation of the evidence is
that genetic variation represented (or reflected) by
rs9939609 amplifies the obesity-inducing effect of FTO,
thereby indirectly affecting susceptibility to infarction.
However, an analysis of the data of
Section~\ref{Data analysis}, based on the method we propose in
Section~\ref{Estimation from matched case-control studies},
shows that the effect of rs9939609 on infarction is
not entirely mediated by body mass. This finding
points to a different theory of the role
of rs9939609 in the development of an infarction.

\newpar Figure 1{\em a} shows a causal diagram representation
of the problem. Causal diagrams are reviewed in Appendix 1.
In the diagram, we let $GENO$ denote genetic variation
responsible for changes in risk of infarction
and correlated with rs9939609. We let $MI$
denote occurrence or nonoccurrence of infarction.
Let $DEMO$ represent the following set of variables:
sex, geographical area of origin and profession. Let $BMI$
represent the body mass index. Let $BEHAVE$ represent
frequent physical exercise and drinking habit.
According to the diagram, the correlation
between $BEHAVE$ and $MI$
is taken to be, in part, induced by shared genetic
or environmental factors, $UNOBSERVED$. The missing
$UNOBSERVED \rightarrow BMI$ arrow represents the assumption
that, conditionally on $BEHAVE$ and $DEMO$, no unobserved
risk factors for obesity are associated with
infarction. Application of the proposed method
to the data of Section~\ref{Data analysis},
under the assumptions of Figure 1{\em a},
shows that the causal effect of $GENO$ on $MI$
is not entirely mediated by $BMI$, in the
sense that a (hypothetical) intervention that fixes
the value of $BMI$ would not completely
block the effect exerted on $MI$ by a (hypothetical)
intervention on $GENO$.
This finding points to new hypotheses about
the role of rs9939609 in susceptibility
to MI. At the end of this paper we discuss the
biological implications of this
finding in the light
of recent experimental research evidence.

\begin{figure}[!ht]
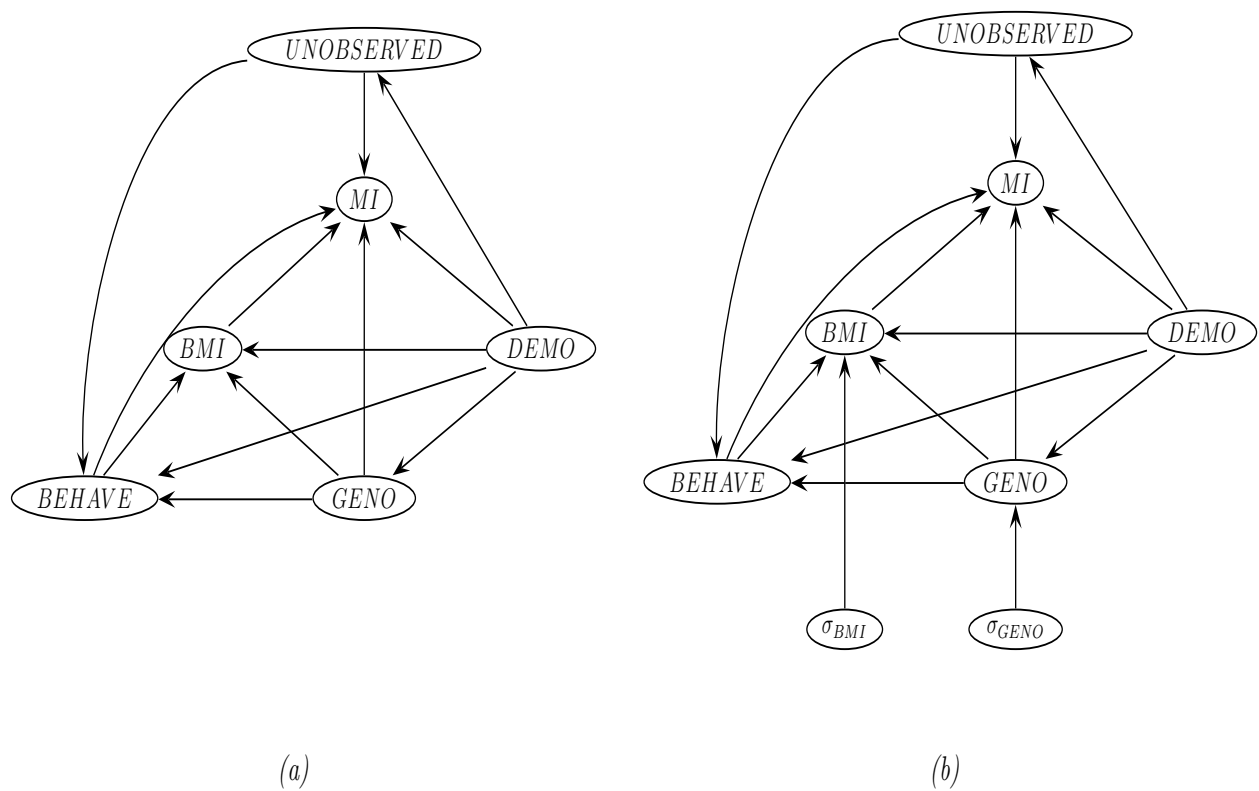

 \begin{center}
\scalebox{0.63}[0.9]{\psframebox[fillstyle=none,fillcolor=white,
   linestyle=none,framesep=.4,arrowsize=2]{%
     \begin{psmatrix}[ref=t,rowsep=1.5cm,colsep=0.1cm]
       &&\Show{$UNOBSERVED$}&\\
        &&\Show{$MI$}\\
       &\Show{$BMI$}&&\Show{$DEMO$}\\
       \Show{$BEHAVE$}&&\Show{$GENO$}&\\
       \\
&&{\large \em (a)}\rule{3cm}{0cm}&
       \ncLine{->,linestyle=solid,arrowsize=4pt 4}{4,3}{2,3}
      \ncLine{->,linestyle=solid,arrowsize=4pt 4}{4,1}{3,2}
       \ncLine{->,linestyle=solid,arrowsize=4pt 4}{3,2}{2,3}
       \ncLine{->,linestyle=solid,arrowsize=4pt 4}{4,3}{3,2}
       \ncarc[arcangle=25,fillstyle=none,arrowsize=4pt 4]{4,1}{2,3}
\ncarc[arcangle=-9,fillstyle=none,arrowsize=4pt 4]{2,3}{2,2}
\ncarc[arcangle=23,fillstyle=none,arrowsize=4pt 4]{4,1}{2,2}
\ncLine{->,linestyle=solid,arrowsize=4pt 4}{1,3}{2,3}
\ncarc[arcangle=-45,fillstyle=none,arrowsize=4pt 4]{1,3}{4,1}
\ncLine{->,linestyle=solid,arrowsize=4pt 4}{4,3}{4,1}
      \ncLine{->,linestyle=solid,arrowsize=4pt 4}{3,4}{1,3}
      \ncLine{->,linestyle=solid,arrowsize=4pt 4}{3,4}{2,2}
      \ncLine{->,linestyle=solid,arrowsize=4pt 4}{3,4}{2,3}
             \ncLine{->,linestyle=solid,arrowsize=4pt 4}{3,4}{3,2}
                    \ncLine{->,linestyle=solid,arrowsize=4pt 4}{3,4}{4,1}
                           \ncLine{->,linestyle=solid,arrowsize=4pt 4}{3,4}{4,3}
     \end{psmatrix}} \psframebox[fillstyle=none,fillcolor=white,
   linestyle=none,framesep=.4,arrowsize=2]{%
     \begin{psmatrix}[ref=t,rowsep=1.5cm,colsep=0.3cm]
       &&\Show{$UNOBSERVED$}&\\
        &&\Show{$MI$}\\
       &\Show{$BMI$}&&\Show{$DEMO$}\\
       \Show{$BEHAVE$}&&\Show{$GENO$}&\\
       &\Show{$\sigma_{BMI}$}&\Show{$\sigma_{GENO}$}&\\
&&{\large \em (b)}\rule{3cm}{0cm}&
       \ncLine{->,linestyle=solid,arrowsize=4pt 4}{4,3}{2,3}
      \ncLine{->,linestyle=solid,arrowsize=4pt 4}{4,1}{3,2}
       \ncLine{->,linestyle=solid,arrowsize=4pt 4}{3,2}{2,3}
       \ncLine{->,linestyle=solid,arrowsize=4pt 4}{4,3}{3,2}
       \ncarc[arcangle=25,fillstyle=none,arrowsize=4pt 4]{4,1}{2,3}
\ncarc[arcangle=-9,fillstyle=none,arrowsize=4pt 4]{2,3}{2,2}
\ncarc[arcangle=23,fillstyle=none,arrowsize=4pt 4]{4,1}{2,2}
\ncLine{->,linestyle=solid,arrowsize=4pt 4}{1,3}{2,3}
\ncarc[arcangle=-45,fillstyle=none,arrowsize=4pt 4]{1,3}{4,1}
\ncLine{->,linestyle=solid,arrowsize=4pt 4}{4,3}{4,1}
      \ncLine{->,linestyle=solid,arrowsize=4pt 4}{3,4}{1,3}
      \ncLine{->,linestyle=solid,arrowsize=4pt 4}{3,4}{2,2}
      \ncLine{->,linestyle=solid,arrowsize=4pt 4}{3,4}{2,3}
             \ncLine{->,linestyle=solid,arrowsize=4pt 4}{3,4}{3,2}
                    \ncLine{->,linestyle=solid,arrowsize=4pt 4}{3,4}{4,1}
                           \ncLine{->,linestyle=solid,arrowsize=4pt 4}{3,4}{4,3}
                                 \ncLine{->,linestyle=solid,arrowsize=4pt 4}{5,2}{3,2}
             \ncLine{->,linestyle=solid,arrowsize=4pt 4}{5,3}{4,3}
     \end{psmatrix}}
     }
     \end{center}
     \vspace{-0.7cm}
 \caption{\small {\em (a)} Causal diagram for our motivating study,
{\em (b)}  same diagram, augmented with intervention
indicators, as explained in Section~\ref{Controlled direct effects}.
 \label{figure1}}
\end{figure}

\section{\textcolor[rgb]{0.00,0.00,0.50}{Controlled direct effects}}
\label{Controlled direct effects}

More in general, let $X$ denote genetic variation of
interest, and the binary variable $Y$ indicate
occurrence ($Y=1$) or non-occurrence ($Y=0$) of the
disease. Let $M$ denote a set of variables
along the causal path from $X$ to $Y$. Define the
{\em direct effect  of $X$ on $Y$, controlled for $M$},
to be the effect exerted on $Y$ by
an {\em intervention} that changes the value of $X$ from
some reference value $x_0$ to $x_1$, while keeping $M$
fixed at some reference value, $m_0$~\cite{Robins1992,
pearldirect}. To formalize this concept, we need to
represent the idea of ``intervention''. This means
to distinguish between
the ``observational'' distribution of the data we
are analyzing, $P_{\emptyset}$,
and the distribution, $P_{xm}$, of the data
we would have obtained had we fixed $X$ to some value $x$
and/or $M$ to some
value $m$. Following Dawid~\cite{Dawid2002},
we label these different distributions by
an {\em intervention indicator} $\sigma_{X}$
and an intervention indicator $\sigma_{M}$, where,
for $H \in (X,M)$, the symbol $\sigma_{H}=\emptyset$
indicates that the value $H$ is observed passively,
and the symbol $\sigma_{H}=h$
indicates that $H$ is set to $h$ by
an intervention. Thus, for a generic variable $W$,
the symbol $P( Y=1 \mid \sigma_X = x, \sigma_{M}=m, W=w)$
denotes the probability of occurrence
of the outcome event, conditional on observing $W=w$,
when we forcefully set $X$ to $x$ and $M$ to $m$.
The direct effect of $X$ on $Y$, controlled for $M$
and conditional on a generic set $W$ of observed variables,
can now be measured in terms of the (causal
conditional) relative risk
\begin{equation}
\label{rr}\frac{P(Y=1 \mid \sigma_X=x_1, \sigma_M=m_0, W)}
{P(Y=1 \mid \sigma_X=x_0, \sigma_M=m_0, W)},
\end{equation}
\noindent or in terms of the (causal conditional) odds ratio
\begin{equation}
\frac{\mbox{odds}(Y=1 \mid \sigma_X=x_1, \sigma_M=m_0, W)}
{\mbox{odds}(Y=1 \mid \sigma_X=x_0, \sigma_M=m_0, W)},
\label{odd aalen}
\end{equation}
\noindent where $\mbox{odds}( Y=1 \mid \sigma_X = x,
\sigma_{M}=m, W=w)=P( Y=1 \mid \sigma_X = x,
\sigma_{M}=m, W=w)/P( Y=0 \mid \sigma_X = x,
\sigma_{M}=m, W=w)$.

\newpar Because our data are generated from $P_{\emptyset}$,
{\em i.e.}, conditional
on $\sigma_X=\emptyset, \sigma_M=\emptyset$,
they will -- in general -- be
uninformative about the interventional probabilities
involved in the direct effect of interest, be it
in the form~\eqref{rr} or in the form~\eqref{odd aalen}.
Does this mean we can {\em never} estimate a direct
effect on the basis of observational data? Luckily, no.
Estimation is possible in special situations, under
identifiability conditions studied in the next  section.
As we shall see, these conditions
can be expressed through the language of conditional
independence~\cite{Dawid1979}, extended by Dawid
to accommodate intervention indicators~\cite{DawidBook1}.
An important tool, in our subsequent discussion,
are causal diagrams extended (augmented) to incorporate
intervention indicators in the form of
additional nodes sending arrows into
their corresponding variables, as in~\cite{DawidBook1}.
One example is the causal diagram of
Figure 1{\em b}, which extends the diagram
of Figure 1{\em a} by adding nodes to represent the
intervention indicators for variables $X$ and $M$.

\vspace{0.5cm}

\begin{figure}[!ht]
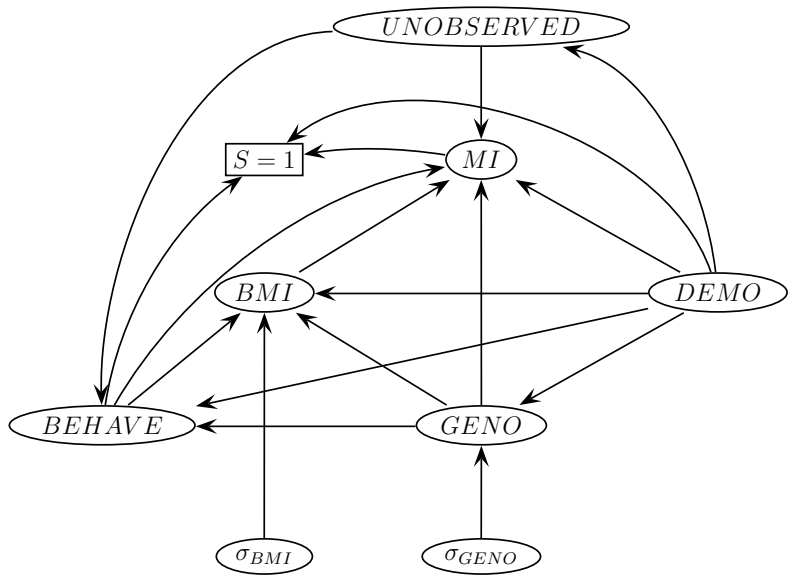

 \begin{center}
\scalebox{0.8}{\psframebox[fillstyle=none,fillcolor=white,
   linestyle=none,framesep=.4,arrowsize=2]{%
     \begin{psmatrix}[ref=t,rowsep=1.5cm,colsep=0.3cm]
       &&\Show{$UNOBSERVED$}&\\
        &\Square{$S=1$}&\Show{$MI$}\\
       &\Show{$BMI$}&&\Show{$DEMO$}\\
       \Show{$BEHAVE$}&&\Show{$GENO$}&\\
       &\Show{$\sigma_{BMI}$}&\Show{$\sigma_{GENO}$}&
       \ncLine{->,linestyle=solid,arrowsize=4pt 4}{4,3}{2,3}
      \ncLine{->,linestyle=solid,arrowsize=4pt 4}{4,1}{3,2}
       \ncLine{->,linestyle=solid,arrowsize=4pt 4}{3,2}{2,3}
       \ncLine{->,linestyle=solid,arrowsize=4pt 4}{4,3}{3,2}
       \ncarc[arcangle=25,fillstyle=none,arrowsize=4pt 4]{4,1}{2,3}
\ncarc[arcangle=-9,fillstyle=none,arrowsize=4pt 4]{2,3}{2,2}
\ncarc[arcangle=23,fillstyle=none,arrowsize=4pt 4]{4,1}{2,2}
\ncLine{->,linestyle=solid,arrowsize=4pt 4}{1,3}{2,3}
\ncarc[arcangle=-45,fillstyle=none,arrowsize=4pt 4]{1,3}{4,1}
\ncarc[arcangle=-55,fillstyle=none,arrowsize=4pt 4]{3,4}{2,2}
\ncarc[arcangle=-35,fillstyle=none,arrowsize=4pt 4]{3,4}{1,3}
\ncLine{->,linestyle=solid,arrowsize=4pt 4}{4,3}{4,1}
      \ncLine{->,linestyle=solid,arrowsize=4pt 4}{3,4}{2,3}
             \ncLine{->,linestyle=solid,arrowsize=4pt 4}{3,4}{3,2}
                    \ncLine{->,linestyle=solid,arrowsize=4pt 4}{3,4}{4,1}
                           \ncLine{->,linestyle=solid,arrowsize=4pt 4}{3,4}{4,3}
                                 \ncLine{->,linestyle=solid,arrowsize=4pt 4}{5,2}{3,2}
             \ncLine{->,linestyle=solid,arrowsize=4pt 4}{5,3}{4,3}
     \end{psmatrix}}
     }
     \end{center}
     \vspace{-0.5cm}
 \caption{\small This diagram has been obtained from Figure 1{\em b}
 by adding the selection indicator node, $S=1$, as explained in
 Section~\ref{Estimation from matched case-control studies}.
 In the diagram, this node receives arrows from $MI$, $DEMO$ and
 $BEHAVE$. This represents the assumption
 that the probability of a generic individual of the study cohort
 being sampled depends on $(MI,DEMO,BEHAVE)$ while
 being, conditional on these variables, independent of $GENO$.
 Its dependence on $GENO$ would violate condition~\eqref{collapsible}.
 \label{figure2}}
\end{figure}

\section{\textcolor[rgb]{0.00,0.00,0.50}{Estimation
from random population samples}}
\label{Estimation from random population samples}

Important results on the
identifiability of controlled direct effects have been
obtained by Robins, Greenland, Didelez, Dawid, Geneletti
and Pearl~\cite{Robins1999,
didelez06,pearldirect,Robins1992}
under the assumption that the population
sample is random. These results
are now summarized, with the
involved assumptions expressed
in the form of conditional
independence conditions between problem variables.

\newpar If there exists a (possibly empty)
set $W$ of observed variables such that, conditionally on $W$,
there is no unobserved confounding of the
relationship between $(X,M)$ and $Y$,
then the direct effect of $X$ on $Y$, controlling for $M$,
is identifiable from random population samples and estimable
via standard regression of $Y$ on $X,M$ and $W$.
The stated condition is equivalent to asking
that $W$ is not a descendant of either $M$ or $X$, and that
the distribution
of $Y$ given $(X,M,W)$ is
the same, regardless of the way the values of
$X$ and $M$ are generated, be it observationally
or by forceful intervention, formally:
\begin{eqnarray}
\label{c0a}
W &\indep &(\sigma_X,\sigma_M),\\
\label{c0b}
Y &\indep &(\sigma_X,\sigma_M) \mid (X,M,W).
\end{eqnarray}
\noindent In fact, it follows from ~\eqref{c0b} that
\[P(Y=1 \mid \sigma_X=x, \sigma_M=m, W)= P(Y=1 \mid X=x, M=m, W),\]
where the righthand side
can be obtained as the fitted
value from a (logistic) regression model, and hence an estimate of the causal
conditional relative risk~\eqref{rr} can also be obtained.
Conditions~(\ref{c0a}-\ref{c0b})
can be checked on an augmented causal diagram by
the $d$-separation criterion~\cite{geiger90,Shpitser2011}
reviewed in Appendix 1, or the equivalent
moralisation criterion~\cite{lauritzen90}.

\small

\newpar \begin{description}
\item[\rm EXAMPLE 1:] the diagram of Figure 1{\em b}
contains causal paths from {\footnotesize $GENO$}
to {\footnotesize $MI$} that do not involve {\footnotesize $BMI$}.
It makes thus sense to test for a direct effect of {\footnotesize
$GENO$} on {\footnotesize $MI$}, controlling for {\footnotesize
$BMI$}. Conditions~(\ref{c0a}-\ref{c0b}) for
this test to be approachable via standard regression
imply the existence of a (possibly empty) set of variables $W$
that satisfies:
\begin{eqnarray}
\label{c0aex}
W &\indep &(\sigma_{GENO},\sigma_{BMI}),\\
\label{c0bex}
MI &\indep &(\sigma_{GENO},\sigma_{BMI}) \mid (GENO,BMI,W).
\end{eqnarray}
\noindent In order
to satisfy~\eqref{c0aex}, the set $W$ must not
contain a member of $BEHAVE$. But then,
because $MI$ and $\sigma_{BMI}$ are $d$-connected
when $BMI$ is in the conditioning set and
$BEHAVE$ is not (in accord with the theory
of Appendix 1), condition~\eqref{c0bex}
will be inevitably violated,
and we conclude that, in this example, the direct
effect of interest cannot be estimated by using standard
regression.
\end{description}

\normalsize

\vspace{0.2cm}

\newpar Estimation of the direct effect of $X$ on $Y$,
controlled for $M$, from prospective observational
data, is possible under more lenient
conditions than~(\ref{c0a}-\ref{c0b}),
although this will occasionally require standard regression
to be abandoned in
favour of the more general method
of $G$--computation~\cite{Robins1992}.
These more lenient conditions require that there be
 a (possibly empty) set
$W$ of non-causal successors of $X$ such that, conditional
on $W$, there is no confounding between $Y$ and
$X$, and a (possibly empty) set $Z$ of non-causal
successors of $M$ such that, conditional
on $(X,Z,W)$, there is no confounding between $Y$
and $M$. All this is formally expressed by
the following conditions:
\begin{align}
 \label{c1}
   &\quad\quad\quad\quad W \indep \; \sigma_X,\\
   \label{c2}
   &\quad\quad\quad\quad Z \indep  \; \sigma_M ,\\
 \label{c3}
   &\quad\quad\quad\quad Y \indep  \; \sigma_X \mid (X,W),\\
   \label{c4}
   &\quad\quad\quad\quad Y \indep  \; \sigma_M \mid (X,M,Z,W),
 \end{align}
\noindent which are similar to those
given in~\cite{didelez06}. Various authors
have discussed $G$-computation~\cite{didelez06,
pearldirect,pearlrob95,Robins1992,Robins1999,tian2010}
or $G$-estimation~\cite{Vansteelandt2009,Vansteelandt2009-genepi,Joffe2009}
of controlled direct effects from a random
population sample in such settings. These authors
use assumptions~\eqref{c1}-\eqref{c4}, although they
sometimes adopt a different "language" to
express them.

\small

\newpar \begin{description}
\item[\rm EXAMPLE 2:] with reference
to the
causal diagram of Figure 1{\em b},
if we specify
{\footnotesize $W \equiv DEMO$}
and {\footnotesize $Z \equiv BEHAVE$}, then
conditions~\eqref{c1}-\eqref{c4} can be written as:
\begin{align}
 \label{c1ex}
   &DEMO &\indep \;\; &\sigma_{GENO},\\
   \label{c2ex}
   &BEHAVE &\indep \;\; &\sigma_{BMI},\\
 \label{c3ex}
   &MI &\indep \;\; &\sigma_{GENO} \;\; &\mid \;\; &(DEMO,GENO),\\
   \label{c4ex}
   &MI &\indep \;\;  &\sigma_{BMI} \;\; &\mid \;\; &(GENO,BMI,BEHAVE,DEMO),
 \end{align}
\noindent Conditions~\eqref{c1ex}-\eqref{c2ex} are
satisfied because neither {\footnotesize $DEMO$} is
a descendant
of {\footnotesize $GENO$}, nor {\footnotesize $BEHAVE$}
a descendant of {\footnotesize $BMI$}.
Condition~\eqref{c3ex} is satisfied because $MI$ and $\sigma_{GENO}$ are
$d$-separated in Figure 1{\em b} when $DEMO$ is in the conditioning set.
Finally, condition~\eqref{c4ex} is satisfied because,
as shown in Appendix 1, nodes $MI$ and $\sigma_{BMI}$
are $d$-separated in Figure 1{\em b} if $BMI$ and $BEHAVE$ are in the
conditioning set. We conclude that the direct effect of
$GENO$ on $MI$, controlling for $BMI$,
is estimable by $G$-computation from
prospective observational data,
under the assumptions of Figure 1{\em b}.
\end{description}

\normalsize

\newpar Conditions~\eqref{c1}-\eqref{c4}
do not prevent $Z$ from being
a descendant of $X$, in which case
the conditioning
on $Z$ will -- in a general prospective study -- create
a spurious association between $X$ and $Y$, even
in absence of the direct effect we wish
to assess~\cite{pearl1998,cole2002}. This
''collider-stratification bias'' will prevent standard regression,
but not necessarily $G$-computation or $G$-estimation, from correctly
estimating the direct effect of $X$ on $Y$,
controlling for $M$, as shown
in~\cite{Vansteelandt2009-genepi}.

\section{\textcolor[rgb]{0.00,0.00,0.50}{Estimation
from matched case-control studies}}
\label{Estimation from matched case-control studies}

Let us now shift attention to the estimation of controlled direct effects
in the context of a {\em retrospective design}. This is, even under
the general conditions~\eqref{c1}-\eqref{c4}, a complicated
task, one reason being the possible ("exposure-induced
mediator-outcome") confounding induced by statistical
dependence between $Z$ and $\sigma_X$ (quite possible
under~\eqref{c1}-\eqref{c4}). The literature on estimating
controlled direct effects from retrospective designs in presence
of this type of confounding is, to the best of our knowledge,
very limited so far. $G$-estimation approaches to
this problem in the context of unmatched case-control studies
have been suggested by Vansteelandt in~\cite{
Vansteelandt2009} and~\cite{Vansteelandt2010}. The
latter paper uses $G$-estimation in combination with
logistic regression. In this section, we shall present
an approach to the problem that works with
{\em matched} case-control studies.

\newpar We start by including in the causal diagram
a special node $S$,
called the {\em selection indicator}, to account for
the non-random sampling involved in case-control studies.
This is exemplified in
Figure 2{\em b}.
The value $S=1$ indicates that the individual
has been selected from the underlying study cohort
for inclusion in the study,
as in~\cite{Hernan2004,Geneletti2009}.
Implicit in a case-control study is the fact that
the selection event, $S$, depends
on the outcome, $Y$, and this is why we
have the $Y \rightarrow S$ arrow in the
diagram. Data analysis is (by tautology)
performed conditional on $S=1$.
Suppose that the usual "rare disease assumption" is valid,
and that the ``collapsibility'' condition
\begin{equation}
X \indep S \mid (Y,M,W),
\label{collapsible}
\end{equation}
\noindent is satisfied, which makes sure
the conditional odds ratio odds$(Y=1 \mid X=x, M=m, W)$
is not affected by the retrospective
sampling~\cite{Didelez2010,whittemore}.
In those situations where the above condition
is satisfied together with~\eqref{c0a}-\eqref{c0b}, a standard
regression approach to the case-control study will
work (conditional logistic regression being one option
when hevcase-control study is matched). In the following,
we are concerned with the more difficult situation
of a matched case-control study where
condition~\eqref{collapsible},
but not ~\eqref{c0a}-\eqref{c0b}, hold.

\newpar Hence suppose that cases
and controls have been 1-to-1 matched with respect to
a set $W$ of variables that satisfies conditions~\eqref{c1}-\eqref{c4}.
Let the $W$-matched pairs be indexed by $i$ (with $i=1,
\ldots , n$) and let the generic notation
$G^{(ij)}$ denote the value of a variable of interest,
$G$, for the $j$th member of pair $i$. Assume the event $Y=1$
is rare (which is often a main motivation for the choice
of a retrospective design), and that the following
model is true:
\begin{align}
\frac{E(Y \mid \sigma_X=x, \sigma_M=m, W, Z)}
{E(Y \mid \sigma_X=0, \sigma_M=0, W, Z)}
&=
\exp(\psi x + \gamma m),
\label{modello}
\end{align}
where expectations $E(.)$ refer to the population distribution.
Then we show in Appendix 2 that the data
will approximately satisfy:
\begin{align}
E^*\left\{(X^{(i1)}-X^{(i0)})\exp(-\psi X^{(i1)}
-\gamma M^{(i1)})\right\}=0,
\label{expectation}
\end{align}
\noindent where the expectation $E^*(.)$ refers to
the observed data distribution under retrospective
sampling. The idea is then to fit the
logistic regression model:
\begin{align*}
{\rm logit} \; P(Y^{(ij)}=1 \mid X^{(ij)}=x, Z^{(ij)}=
z, M^{(ij)}=m) &=
\alpha + \delta x + \beta z +\eta m + b^{(i)},
\end{align*}
\noindent where $b^{(i)}$ is a mean zero
random effect, which expresses the contribution
for matched pair $i$. A maximum likelihood
estimate of the remaining parameters,
$(\alpha,\delta,\beta,\eta)$, can be obtained via
conditional logistic regression, for example
by using the {\tt CLOGIT} procedure
in {\tt R}. Under the
"no confounding" conditions~\eqref{c2} and \eqref{c4}, the estimate
of $\eta$, denoted by
$\hat{\eta}$, encodes the conditional causal
effect of $M$ on $Y$, represented
in Equation~\eqref{modello} by
the symbol $\gamma$.
Equation~\eqref{expectation} then
justifies the use of the following
conditional score equation:
\begin{align}
0 &= \sum_{i=1}^n
(x^{(i1)}-x^{(i0)}) \; {\rm exp} \left( -\psi x^{(i1)}
-\hat{\eta} m^{(i1)} \right)
\label{conditional score}
\end{align}
\noindent for estimating the direct
effect of interest, which is encoded by $\psi$.
An estimator
for the variance of $\hat{\psi}$ is derived in
the last paragraph of Appendix 2.

\small

\newpar \begin{description}
\item[\rm EXAMPLE 3:] it is easy to show,
along the lines of
Example 2, that, for {\footnotesize $W \equiv DEMO$}
 and {\footnotesize $Z \equiv BEHAVE$}, the
causal diagram of Figure 2
satisfies
conditions~\eqref{c1}-\eqref{c4}
for {\footnotesize $W \equiv DEMO$} and
{\footnotesize $Z \equiv BEHAVE$},
and the collapsibility condition
{\footnotesize $GENO \indep S \mid (MI, BMI, DEMO)$},
as well. Because of the above considerations,
and because
early infarction is a rare disease,
we conclude that the direct effect of
{\footnotesize $GENO$}
on {\footnotesize $MI$}, controlling for {\footnotesize $BMI$},
is estimable by $G$-computation from
matched case-control data,
under the assumptions of Figure 2.
\end{description}

\normalsize

\section{\textcolor[rgb]{0.00,0.00,0.50}{Back
to our motivating study}}
\label{Data analysis}

Within an Italian study in the genetics of
infarction~\cite{Ardissino2011},
cases were ascertained
on the basis of hospitalization for
acute myocardial infarction between ages
40 and 45, during the
1996 -- 2002 period. This study involves the
variables represented in Figures 1 and 2, and
which we continue to denote through the
symbols introduced in Section~\ref{Motivating study}.
The controls were selected by
matching them to the cases over sex,
geographical area of origin
and profession ($DEMO$).

\newpar Our aim here is to estimate,
on the basis of the study data, the
effect of genetic variation reflected
by rs9939609 ($GENO$) on risk of early infarction ($MI$),
controlling for body mass ($BMI$).
We work under the assumptions represented
in the diagram of Figure 2, which appear legitimate,
especially when one considers
the narrow range of ages at infarction
represented in our sample of cases.
Under such assumptions, we have already seen
in Example  3 that the direct effect of
interest is estimable
by using the algorithm described in the previous section.

\newpar The distribution of the rs9939609
genotype in sample cases and controls
is summarized in Table~\ref{table1}.
No major departure
from Hardy-Weinberg equilibrium in controls
was detected.

\begin{table}[ht]
\begin{center}
\begin{tabular}{r|rr}
 \hline
number of copies of the\\
major rs9939609 allele& controls & cases \\
 \hline
 0 & 305 & 380 \\
 1 & 889 & 921 \\
 2 & 644 & 537 \\
  \hline
\end{tabular}
\end{center}
\caption{\small Distribution of the rs9939609
genotype in sample cases and controls.\\
\label{table1}}
\end{table}

\begin{table}[ht]
\begin{center}
\begin{tabular}{r|rrc}
 \hline
&OR & $p$-value & $95\%$ confidence interval\\
 \hline
rs9939609 wild-type homozygosity? & 0.76 & 0.0001 &
0.65   -  0.87\\
  \hline
\end{tabular}
\end{center}
\caption{\small Results from the fitting of a conditional
logistic model for the dependence of occurrence
of early myocardial infarction on rare rs9939609
homozygosity, without any adjustment for
other variables in the model. This produces
an estimate of the {\em total} effect of the
rs9939609 rare homozygosity on susceptibility
to early myocardial infarction, on an
odds ratio of disease scale, reported
in the OR column of the table.\label{table2}}
\end{table}

\begin{table}[ht]
\begin{center}
\begin{tabular}{r|rrc}
 \hline
& OR &$p$-value& $95\%$ confidence interval\\
 \hline
rs9939609 wild-type homozygosity? & 0.81 & 0.007 &0.7   -  0.94\\
 body mass index & 1.15 & $<$ 2e-16 &1.12   - 1.17\\
  \hline
\end{tabular}
\end{center}
\caption{\small Results from the fitting of a conditional
logistic model for the dependence of occurrence
of early myocardial infarction on rare rs9939609
homozygosity, adjusting for
body mass index.
\label{table3}}
\end{table}

\begin{table}[ht]
\begin{center}
\begin{tabular}{rrrc}
 \hline
&OR & $p$-value & $95\%$ confidence interval\\
 \hline
rs9939609 wild-type homozygosity? & 0.84 & 0.02 &
  0.72  -  0.98\\
body mass index & 1.14 & $<$ 2e-16 &
 1.11  -  1.16\\
occasional physical exercise? & 0.61 &  1.41e-07 &
 0.50  -  0.73\\
frequent physical exercise? & 0.53 & 3.13e-13&
  0.44  -  0.63\\
drinking habit? & 1.36 & 7.48e-05 &
 1.17  -  1.59\\
  \hline
\end{tabular}
\end{center}
\caption{\small Results from the fitting of a conditional
logistic model for the dependence of occurrence
of early myocardial infarction on rare rs9939609
homozygosity, adjusting for body mass index,
physical exercise and drinking habit.\label{table4}}
\end{table}

\newpar Table~\ref{table2} summarizes results from
the fitting of a conditional
logistic model for the dependence of occurrence
of early myocardial infarction on wild-type rs9939609
homozygosity, without any adjustment for
other variables in the model (except, of course,
for the matching variables). This yielded
an estimate of 0.76 for the
{\em total} effect of
rs9939609 rare homozygosity on infarction, on
an odds ratio scale, which is
significantly different from the null at a 0.0001
level of significance. This can be interpreted
as evidence of an ``overall protective'' effect of
the major rs9939609 allele.

\newpar When we further included
body mass as an additional explanatory variable
in the model, we obtained the results
of Table~\ref{table3}, where the
effect of rs9939609 wild-type homozygosity
on infarction, 0.81 on an odds ratio scale,
significantly departs from the null
at a 0.007 level of significance.
Unfortunately, because conditions~(\ref{c0a}-\ref{c0b})
are violated
by the diagram of Figure 2, we cannot take this estimate
as a valid measure of the direct effect of rs9939609 wild-type homozygosity
on infarction, controlling for body mass. One problem here is,
in fact, that physical exercise and drinking
are potential confounders of the association between
body mass and myocardial infarction.

\newpar Can this problem be overcome by
including the $BEHAVE$ variables
-- physical exercise
and drinking habit -- as additional
covariates in the regression model? When we did so,
the estimated effect of rs9939609 wild-type
homozygosity on infarction was 0.84, which
is a significant (at a 0.02 level) departure from the null (see Table 4). Again, because the
causal diagram of Figure 2 violates conditions~\eqref{c0a}-\eqref{c0b},
our method does not guarantee that the above estimate, obtained by
standard regression, is a valid measure of the direct
effect of interest. One problem being that
the conditioning on $BEHAVE$ opens the
$GENO \rightarrow BEHAVE \leftarrow U \rightarrow MI$
path (see Appendix 1) and, as a consequence, it introduces
a spurious, non causal, association between $GENO$ and $MI$,
so called collider-stratification bias. We must accept the
fact that, according to our method, no valid
estimate of the direct effect of interest can be
obtained by standard regression. Luckily, because the
causal diagram of Figure 2 satisfies
conditions~(\ref{c1}-\ref{c4}, \ref{collapsible}),
our method tells us that a valid estimate of the
direct effect of rs9939609 on infarction, controlling
for body mass, can be obtained by using the
$G$-estimation procedure of the preceding section. This
yields an estimate of $0.72$, with a $95\%$ confidence interval
of $(0.62, 0.84)$, on a relative risk scale.
This estimate differs appreciably from the estimates
obtained in previous steps of the analysis. The fact
that the latter estimate refers to the relative risk scale,
rather than the odds ratio scale, does
not entirely explain this difference in view
of the low prevalence of early-onset myocardial infarction.

\newpar From a substantive point of view, our finding suggests
that genetic variation represented
by rs9939609 may influence heart
disease via pathways
different from those involved in body mass.
A biological interpretation of this
finding is given at the end of
the next section.

\section{\textcolor[rgb]{0.00,0.00,0.50}{Discussion}}

In this paper, we have started by examining conditions
under which controlled direct effects can be estimated
from prospective observational data via standard regression.
When these conditions are violated, the direct effect of
interest is sometimes still estimable from a prospective
study, albeit not via regression. We have examined the
more general conditions under which a controlled direct effect
is estimable via $G$-computation, and we have
expressed them as properties of
a causal diagram representation
of the problem. Then, in consideration of the
increasing importance of matched case-control studies
in genetic epidemiology, we have shifted attention to
this class of studies. We have proposed an algorithm
for the $G$--estimation of controlled direct effects
from matched case-control studies, and characterized the necessary
conditions for algorithm validity in terms of
conditional independence properties of the
causal diagram representation of the problem.

\newpar The proposed method is also relevant
in situations where the notion
of "case" is not the usual one.
Examples are offered by the papers of
Cordell and colleagues~\cite{Cordell2004},
and of Bernardinelli and colleagues~\cite{BernardinelliEtalPlos},
where genetic effects are estimated by conditioning
on parental genotypes, using data from proband-parent trios.
These papers essentially
perform a matched case-control analysis
via conditional logistic regression, using the
case and one or more "pseudocontrols"
derived from the untransmitted parental haplotypes.
This approach could be combined with the methods
presented in this paper to assess {\em direct}
controlled genetic effects.

\newpar In the context of  retrospective designs, further study is
warranted of identification results for controlled direct
effects in specific model classes, as well as for so-called
"natural" direct and indirect effects~\cite{pearl2011}.
In addition, further work is needed to investigate
whether direct effect estimators can be constructed
on the basis of matched case-control studies, which
are either more efficient than the estimator proposed
in this paper,
or less dependent on a rare disease assumption. Finally,
future work will also focus on inference under alternative
strategies for the selection of controls
in a retrospective study.

\newpar We have illustrated the method with the aid
of a study in the genetics of
myocardial infarction. Our analysis detected presence of a direct
effect of rs9939609 on infarction, controlling for body
mass. This finding suggests that the effect of this SNP
on susceptibility to infarction is not totally explained
in terms of a deleterious effect of FTO on body mass. This
finding points to a number of possible hypotheses. Very relevant here
is recent evidence that SNPs can, in general, exert an influence on the
expression of relatively distant (in terms of DNA stretch)
genes. In our case, it could be that rs9939609 drives the
expression of a gene other than FTO,  functionally unrelated
with FTO, whose effect on risk of infarction is not mediated
by body mass. And hence the direct effect.
Such hypothesis is corroborated by biological
evidence that the FTO is located in a genomic region containing
highly conserved genomic regulatory blocks which,
according to a well established theory, are likely
to drive the expression of distant genes~\cite{Kikuta2007,
Bejerano2004,Sandelin2004}. The above considerations have
useful implications with respect to possible experiments
to elucidate the mechanism. It is not unlikely that
rs9939609 may simultaneously drive the expression of different,
and functionally unrelated, genes. Such a multi-effect
pattern could be common. For example a
recent study~\cite{Harismendy2011} shows that
SNPs in the 9p21.3 region of DNA, notoriously associated
with susceptibility to infarction, not only control nearby
genes, but also the expression of the quite distant IFNA21 gene.
Generalizing on this example, one might conjecture
that many SNPs exert their influence on disease
susceptibility through non-overlapping pathways, and that this will,
in many cases, result in evidence of direct and indirect effects
that our method is able to capture.

\section*{\textcolor[rgb]{0.00,0.00,0.50}{Acknowledgments}}

The authors thank Elisabeth Coart for preliminary data analyses,
and Drs. Diego Ardissino and Pierangelica Merlini
for providing their insight into the clinical
problem. The first author acknowledges support from
the UK Medical Research Council Grant no.
G0802320 ({\tt www.mrc.ac.uk}), and by
the Cambridge Statistics Initiative.
The second author acknowledges support from research project
G.0111.12 of the Fund for Scientific Research (Flanders), IAP
research network grant nr. {\tt P06/03} from the Belgian government
(Belgian Science Policy), and Ghent University (Multidisciplinary
Research Partnership ``Bioinformatics: from nucleotides to networks'').

\small

\section*{\textcolor[rgb]{0.00,0.00,0.50}{Appendix 1: Causal diagrams}}

Causal diagrams~\cite{pearlbook,Dawid2002,Greenland2000}
consist of a set of {\em nodes} representing variables in
the problem, and directed {\em arrows} connecting pairs of nodes,
as in Figure 1, for example. The same,
elliptical, shape is used for all nodes. In particular,
no distinction is made, in terms of node shape, between observed
and unobserved variables/nodes, one reason being that this is not a distinction
that has to do with the causal structure of the system under study.
The arrows
represent direct causal influence, in a sense to
be made clear. A {\em path} is a sequence of
distinct nodes where any two adjacent nodes in the sequence are connected by
an arrow. A {\em directed path} from a node $X$ to a node $B$ is a path where all
arrows connecting nodes on the path point away
from $A$ and towards $B$. For example, in
the graph of Figure 2, the sequence

\begin{quote}
\begin{center}
$GENO, \; BEHAVE , \;
BMI ,  \; MI,  \;  UNOBSERVED$
\end{center}
\end{quote}

\noindent is a path between $GENO$ and $UNOBSERVED$,
but not a directed one.

\newpar If $A$ has a directed path to
$B$ then $A$ is an {\em ancestor} of $B$, and $B$ a {\em descendant} of $A$. By convention,
$A$ is both an ancestor and a descendant of $A$. If
an arrow points from $A$ to $B$, then $A$ is called a {\em parent} of $B$. In this paper,
we restrict to causal diagrams
which have the form of a directed acyclic graph
(DAG), that is, a directed graph where for any directed path from $A$ to $B$,
node $B$ is not a parent of $A$. A probability distribution
over the set of nodes of the graph is said
to be {\em Markov} with respect
to the graph if it can be expressed as a product
of factors, where each factor
is the conditional probability of a node of the graph, given its parents in the graph.

\newpar A consecutive triple of nodes, $A,B,C$ say, on a path is called a
{\em collider} if the arrow between $A$ and $B$ and the arrow between $C$ and
$B$ both have arrowheads pointing to $B$.  For example, in Figure 1,
node $BEHAVE$ is a collider on the

\begin{quote}
\begin{center}
$GENO \rightarrow BEHAVE \leftarrow UNOBSERVED$
\end{center}
\end{quote}

\noindent path. Any other consecutive triple is called
a {\em non-collider}. A path between two nodes,
$A$ and $B$ say, is said to be {\em blocked} by a
set $C$ if either for some non-collider on the path,
the middle node is in $C$, or if the  path contains
a collider such that no descendant of  the middle
node of such collider is in $C$.  For example, in
the graph of Figure 2, the path

\begin{quote}
\begin{center}
$GENO \rightarrow BEHAVE \rightarrow
BMI \rightarrow MI \leftarrow UNOBSERVED$
\end{center}
\end{quote}

\noindent is blocked by any set of nodes that contains either
or both of $(BEHAVE,BMI)$, and/or does not contain $S$ or $MI$.
In particular, the path is blocked by the empty set of
nodes.

\newpar For disjoint sets $A,B,C$ of nodes in a DAG
we say $A$  is $d$-separated from $B$ given $C$ if
every path from a node in $A$ to a node in $B$ is
blocked by $C$. If $A$ is not $d$-separated from $B$ given
$C$, we say $A$ is $d$-connected to $B$ given $C$.
For example, in the diagram of Figure 1{\em b}, nodes $\sigma_{BMI}$ and $MI$
are $d$-separated given $(GENO, BMI, BEHAVE, DEMO)$. This is
because all paths between $\sigma_{BMI}$ and $MI$ contain
at least one of the following
non-colliders:
\begin{center}
$(MI,BEHAVE,BMI)$,
$(UNOBSERVED,BEHAVE,BMI)$,
$(UNOBSERVED,DEMO,GENO)$,
$(UNOBSERVED,DEMO,BMI)$,
$(MI,GENO,BMI)$,
\end{center}
\noindent all of which are blocked
by virtue of the fact that $BEHAVE$, $DEMO$ and $GENO$ are in the
conditioning set. As a further example, the reader is invited
to check that that $\sigma_{GENO}$ and $MI$ are
$d$-connected in the diagram of Figure 1{\em b} if
$BMI$, but not $GENO$, is in the conditioning set.
Two sets of nodes, $A$ and $B$ say, that are $d$-separated
given a third set $C$, are conditionally independent, in
a probabilistic sense, given $C$, under any distribution
that is Markov with respect to the graph. By contrast,
if $A$ and $B$ are $d$-connected given
$C$, there exists some probability
distribution which is Markov with respect to
the graph, under which
$A$ and $B$ are {\em not} conditionally
independent, given $C$.

\section*{\textcolor[rgb]{0.00,0.00,0.50}{Appendix 2}}

We now prove that, under conditions (\ref{c1}-\ref{c4},
\ref{collapsible}),
model~\eqref{modello}
and a matched case-control sampling regime of the
kind described in Section~\ref{Estimation
from matched case-control studies}, the data
approximately satisfy Equation~\eqref{expectation},
which we here repeat for the reader's convenience:
\begin{align}
E^*\left\{(X^{(i1)}-X^{(i0)})\exp(-\psi X^{(i1)}
-\gamma M^{(i1)})\right\}=0,
\label{expectation again}
\end{align}
\noindent where the expectation $E^*(.)$ refers to
the observed data distribution under retrospective
sampling.

\newpar Model~\eqref{modello} implies:
\begin{align*}
\frac{E(Y \mid \sigma_X=x, \sigma_M=m, W, Z)}
{E(Y \mid \sigma_X=x, \sigma_M=0, W, Z)}
&=
\exp(\gamma m),
\end{align*}
\noindent from which
we obtain:
\begin{align*}
E(Y \mid \sigma_X=x, \sigma_M=m, X=x, M=m, W, Z) \;\;
\exp(-\gamma m) &=
E(Y \mid \sigma_X=x, \sigma_M=0, W, Z),
\end{align*}
\noindent because for a generic
variable $H$ the equality $\sigma_H=h$ logically implies $H=h$;
at least, this is true under the so-called consistency assumption
that setting $H$ to $h$ by intervention has no effect amongst
those for whom $H=h$ is naturally observed.

\newpar Thanks to the conditioning on $X=x$ and $M=m$,
we may now bring the $\exp(-\gamma m)$ factor of the left
hand side into the expectation, and further multiply both
sides of the equation by the factor $\exp(-\psi x)$, so as to obtain:
\begin{align*}
E \left[ Y \exp(-\psi x -\gamma m) \right.&\left. \mid \sigma_X=x, \sigma_M=m, X=x, M=m, W, Z \right] =\\
&= E \left[ Y  \exp(-\psi x) \mid \sigma_X=x, \sigma_M=0, W, Z \right].
\end{align*}
\noindent Then, by virtue of conditions
~\eqref{c3}-~\eqref{c4}, respectively, we
can eliminate the conditioning
on $\sigma_X=x$ and $\sigma_{M}=m$ from
the left hand side of the equation,
which leads to:
\begin{align*}
E \left\{ Y \; \exp(-\psi x-\gamma m) \mid X=x, M=m, W, Z \right\}
&= E \left\{Y \exp(-\psi x) \mid \sigma_X=x, \sigma_M=0, W, Z \right\}.
\end{align*}
\noindent where the expectation at the left hand side
is taken with respect to the population
distribution (which is what the absence of the $\sigma$ indicators
in the conditioning part means). From
the above equation, by virtue of~\eqref{modello}, we obtain:
\begin{align}
E \left\{ Y \; \exp(-\psi x-\gamma m) \mid X=x, M=m, W , Z\right\}
&= E \left[ Y \mid \sigma_X=0, \sigma_M=0, W, Z \right].
\label{zoccolo}
\end{align}
\noindent The above equality implies that, conditionally on $W$ and $Z$,
the random variable
$$
Y \; \exp(-\psi X-\gamma M)
$$
\noindent is,
in expectation under the
population distribution, independent of $(X,M)$ and therefore,
in a sample from a random population, the quantity:
\begin{align}
(X_i - E\{ X \}) \; Y_i \; \exp(-\psi X_i - \gamma M_i)
\end{align}
\noindent has, conditionally on $W$ and $Z$, zero mean at the true parameter values.

\newpar Recall that we are dealing with a sample from
a 1-to-1 matched case-control study. For the affected member of the
$i$th matched set, consider the quantity:
\begin{align*}
E^* & \left\{ X^{(i1)}\exp(-\psi X^{(i1)}-\gamma M^{(i1)} )  \mid W=W^{(i1)} \right\} =\\
   &= E \left\{ XY\exp(-\psi X-\gamma M) \mid W=W^{(i1)},Y=1 \right\},\\
   &= E \left\{ XY\exp(-\psi X-\gamma M) \mid W=W^{(i1)} \right\}/ P(Y =1 \mid W=W^{(i1)})
   \end{align*}
   \noindent where the expectations $E(.)$
   are taken with respect to the population distribution. By
   virtue of the above independence property, the above equation
   can be rewritten as:
   \begin{align*}
   &E\left\{ X \mid W=W^{(i1)}\right\} \; E\left\{Y\exp(-\psi X^{(i1)}-\gamma M^{(i1)}) \mid
W=W^{(i1)} \right\}/ P(Y=1 \mid W=W^{(i1)}).
\end{align*}
\noindent which, in the light of~\eqref{zoccolo}, can be
written as:
\begin{align*}
&= E[X  \mid W=W^{(i1)}] \; E \left[Y \mid \sigma_X=0, \sigma_M=0, W=W^{(i1)}\right] / P(Y=1 \mid W=W^{(i1)}),
\end{align*}
\noindent Further, note that by a similar reasoning
\begin{align*}
E^* & \left[X^{(i0)}\exp(-\psi X^{(i1)}-\gamma M^{(i1)}) \mid W=W^{(i1)} \right]\\
&= E[X \mid W=W^{(i0)},Y=0] \; E\left[Y \exp(-\psi X-\gamma M) \mid W=W^{(i1)},Y=1\right]\\
&= E[ X \mid W=W^{(i0)},Y=0 ] \; E \left[Y \mid \sigma_X=0, \sigma_M=0, W=W^{(i1)}\right]\\
& /P(Y=1 \mid W=W^{(i1)}).
\end{align*}
\noindent Under a rare disease assumption, we have $E[X \mid W,Y=0]
\approx E[X \mid W]$, which gives Equation~\eqref{expectation again}.
{\em Quod erat demonstrandum}.

\newpar In the remaining part
of this Appendix, we derive an estimator
for the variance of the estimate
of the parameter $\psi$ of Equation~\eqref{conditional score}.
We start by
defining
$\theta \equiv (\psi, \delta, \gamma,\beta)$
and let $U_i(\theta)$ be given by:
\begin{align*}
U_i(\theta) &= \left(
\begin{array}{c}
(x^{(i1)}-x^{(i0)}) \; {\rm exp} \left( -\psi x^{(i1)}
-{\eta} m^{(i1)} \right)\\
\left( \begin{array}{c}
 x^{(i1)}-x^{(i0)} \\
 m^{(i1)}-m^{(i0)} \\
 z^{(i1)}-z^{(i0)} \\
\end{array}
\right) \; {\rm expit} \left( -\delta (x^{(i1)}-x^{(i0)})
-\eta (m^{(i1)}-m^{(i0)}) - \beta (z^{(i1)}-z^{(i0)})
\right)
\end{array}
\right).
\end{align*}
\noindent Let $\hat{\theta}$ denote the estimate
of $\theta$ obtained by our method.
The variance of $\hat{\theta}$ is
well approximated in large samples by the following
sandwich estimator:
\begin{align}
\frac{1}{n} \E^{-1} \left(  \frac{\partial U_i(\theta)}{\partial \theta}  \right)
{\rm Var} (U_i(\theta)) \E^{-1} \left(  \frac{\partial
U_i(\theta)}{\partial \theta}  \right)^{T},
\label{matrix}
\end{align}
\noindent where ${\rm Var} (U_i(\theta))$ can be estimated by
calculating $U_i({\theta})$, then taking
the sample variance of these contributions for
all subjects, and finally evaluating at $\hat{\theta}$. The quantity
$\E \left(  \partial U_i(\theta) / \partial \theta  \right)$ can
be estimated by first calculating the gradient matrix
$\partial U_i(\theta) / \partial \theta $ for
each subject, evaluating it at $\hat{\theta}$ and
then calculating the sample average (over
all subjects) of each component of the matrix.
In this gradient matrix, the element in the
$j$th row and $l$th column should be the derivative of the
$j$th component of $U_i(\theta)$
with respect to the $l$th component of $\theta$. The
first diagonal element of the
resulting matrix~\eqref{matrix} gives
the approximate variance of $\hat{\psi}$.

{\small \bibliographystyle{plain}
\bibliography{bibliografianuova} }

\end{document}